\documentclass[nofootinbib,onecolumn]{revtex4}
\usepackage{graphicx}
\usepackage{bm}
\usepackage{amsfonts}
\usepackage{amsmath}
\usepackage{amssymb}
\usepackage{subfig}
\usepackage{hyperref}
\usepackage{verbatim}

\usepackage{hyperref}
\hypersetup{colorlinks=true, linkcolor=blue, citecolor=blue, filecolor=blue, urlcolor=blue,
            pdfproducer={Latex},
            pdfcreator={pdflatex}}

\begin{document}

\title{Dark energy with zero pressure: Accelerated expansion and large scale structure in action-dependent Lagrangian theories}

\date{\today}

\author{Thiago R. P. Caram\^es}
\email{trpcarames@gmail.com}
\affiliation{N\'ucleo Cosmo-ufes \& Departamento de F\'{\i}sica, Universidade Federal do Esp\'irito Santo, \\Avenida Fernando Ferrari 514, Vit\'oria 29075-910, Esp\'irito Santo, Brazil
}
\author{H. Velten}%
\email{velten@pq.cnpq.br}
\affiliation{N\'ucleo Cosmo-ufes \& Departamento de F\'{\i}sica, Universidade Federal do Esp\'irito Santo, \\Avenida Fernando Ferrari 514, Vit\'oria 29075-910, Esp\'irito Santo, Brazil}
\author{J. C. Fabris}%
\email{julio.fabris@cosmo-ufes.org}
\affiliation{N\'ucleo Cosmo-ufes \& Departamento de F\'{\i}sica, Universidade Federal do Esp\'irito Santo, \\Avenida Fernando Ferrari 514, Vit\'oria 29075-910, Esp\'irito Santo, Brazil}
\affiliation{National Research Nuclear University MEPhI \\ Kashirskoe sh. 31, Moscow 115409, Russia}
\author{Matheus J. Lazo}\email{matheusjlazo@gmail.com}
\affiliation{Instituto  de Matem\'atica,  Estat\'istica  e F\'isica – FURG, Rio Grande 96201-900, Rio Grande do Sul, Brazil.}

\begin{abstract}
We develop a cosmological model based on action-dependent Lagrangian theories. The main feature
here is the nonconservation of the energy momentum tensor due to the nontrivial geometrical construction
of the theory. We provide the basic set of equations necessary to study both the cosmological background
expansion as well as the linear matter perturbation growth. We show that the simplest realization of the
Universe as described by only one component is not viable as expected from the existing correspondence
between this model and the case of viscous cosmological fluids. However, modeling the energy content of
the Universe as composed by two pressureless fluids, i.e., one a typical cold dark matter fluid and the other
a pressureless “dark energy” fluid which is responsible for driving the late-time acceleration expansion, is
qualitatively compatible with observational data.   
\end{abstract}


\maketitle
\date{\today}

\section{Introduction}
The real world is pervaded by all sorts of dissipative
processes; what imposes a suitable description of these
phenomena on any consistent physical process is its
respective scope. Nonetheless, this issue is usually left
outside the standard variational formulations. In the
classical mechanics context, one usually approaches it
by using the so-called Rayleigh dissipation function, which
is a useful tool to deal with dissipative forces with linear
dependence in the velocity \cite{rayleigh,goldstein,whittaker}. Although this method
provides a quite handy procedure for the description of
such friction forces, the Rayleigh’s function method pos-
sesses clear limitations. For instance, it fails in addressing a
broader class of dissipating cases existing in nature with
more general dependencies upon the velocity and the
history of the system. Besides, this function arises as a
correction in the Euler-Lagrange equation which does not
affect at all the underlying variational formalism. In fact, it
was demonstrated in \cite{bauer} that the Rayleigh's function is prohibited from emerging from a variational principle,
unless the dissipative coefficient is not a constant anymore.
These limitations are significant and point towards the
search for possible extensions.
Many attempts at incorporating dissipation effects into
the traditional principle of least action were made over
the last century. They basically rely on the use of
time-dependent Lagrangians \cite{timeD}, auxiliary coordinates that describe the reverse-time system \cite{Morse}, or a fractional derivatives formalism \cite{fracD,fracD1}. However, these proposals face
serious conceptual (or operational) obstacles which can
undermine them as feasible alternatives, as they can
either plague the theory with nonphysical Lagrangians or
give rise to nonlocal differential operators, whose implementation introduces an undesirable complexity to the
study of some problems.
Another noteworthy alternative dates back to the
1930s, where G. Herglotz presented an elegant variational
treatment to this issue by assuming action-dependent
Lagrangians in the context of classical mechanics. In his
approach, Herglotz was indeed successful in describing the
class of dissipative systems whose motion is damped by a
friction force, characterized by the aforementioned term
proportional to velocity \cite{herglotz}. Most important, the Herglotz variational formulation is free from the conceptual and practical obstacles found in the others approaches.  In recent
works, one of us, in partnership with some colleagues,
extended the original Herglotz formalism to a covariant
language \cite{lazo,lazo2}. As it was shown in the work \cite{lazo}, such a covariant generalization laid the cornestone for the construction of a new theory of gravity in which dissipative effects would be a natural consequence, coming from first principles and having a purely geometric origin.  In this
vein, the authors derived explicitly, from a generalized
action, the modified field equations of the theory.
Additionally, they showed that the dynamics of the model
shall include a nonstandard conservation law for the
energy-momentum tensor as a consequence of the breaking
of diffeomorphism resulting from the incorporation of
dissipative processes into the description of the gravity.
Some possible effects of this theory of gravity on the
cosmological environment were explored by us in a further
work \cite{fabris}. There we verified an analogy of the background
dynamics arising in this model with that one of a bulk
viscous cosmology in the Eckart formalism. This feature
provided us an immediate mapping between the coefficient
of bulk viscosity with the coupling parameter encoding the
modification of gravity. We also addressed the evolution of
matter perturbations at the linear level, which allowed us to
glimpse a possible way out to avoid some drawbacks faced
by the viscous model, perhaps leading to the reconcilement
of the obtained pattern of perturbations growth with the
expected background dynamics.
In this work, we deepen our previous study on the
cosmological aspects of the action-dependent gravity by
investigating its viability in light of some important obser-
vational data. As a first step, we consider a model endowed
with a single matter fluid whose conservation departs from
the usual one due to the geometriclike dissipative effects
induced by this modified gravity. Due to the inviability of the
single component model shown below, we model in Sec. III a cosmological model in which there are two pressureless
components. One of them remains obeying the conservation
equation while the second couples to the nontrivial geo-
metrical construction of the action-dependent Lagrangian
theory and therefore yields to an accelerated expansion. We
investigate the viability of this model and then present our
conclusion in Sec. IV.  

\section{Theory}

The so-called Herglotz problem, originally built within
classical mechanics scenario, consists in generalizing the
action principle by introducing in the Lagrangian an action-
dependence as follows 
\begin{equation}
S= \int \mathcal{L}(x,\dot{x}, S)dt.
\end{equation}
In his work he demonstrated that such a formulation showed up as a successful way to describe dissipative phenomena from first principles. A covariant generalization of this problem provides a theory of gravity recently found by Lazo {\it et. al} \cite{lazo} in which
\begin{equation}
\label{Lag}
\mathcal{L}=\sqrt{-g}(R-\lambda_{\mu}s^{\mu})+\mathcal{L}_m
\end{equation}
where $s^{\mu}$ is an action-density field and $\lambda_{\mu}$ is a coupling term which may depend on the spacetime coordinate. According to the Lazo {\it et al} approach, this action-dependence introduced in (\ref{Lag}) through $s^{\mu}$ is only with respect to the standard Einstein-Hilbert action, not to the matter action. As such, the modification of gravity provided by this theory is of a purely geometric nature. This approach leads to a geometrical  viscous  gravity  model in which the dynamics of this theory is described by the generalized field equations 
\begin{equation}
\label{Keqs}
R^{\mu}_{\nu}-\frac{1}{2}R \delta^{\mu}_{\nu} + K^{\mu}_{\nu}-\frac{1}{2}K \delta^{\mu}_{\nu} = \frac{8\pi G}{c^4} T^{\mu}_{\nu}
\end{equation}
along with the modified conservation law
\begin{equation}
 \label{consL}
 \frac{8 \pi G}{c^4}T^{\mu}_{\nu;\mu}=K^{\mu}_{\nu;\mu}-\frac{1}{2}K_{;\nu},
\end{equation}
where the semicolon symbol denotes covariant derivatives. The departure from GR is clearly encoded in the quantity $K_{\mu\nu}$ given by
\begin{equation}
K_{\mu\nu}= \lambda_{\alpha}\Gamma^{\alpha}_{\mu\nu}-\frac{1}{2}\left(\lambda_{\mu}\Gamma^{\alpha}_{\nu\alpha}+\lambda_{\nu}\Gamma^{\alpha}_{\mu \alpha}\right),
\end{equation}
where $\lambda_{\mu}$ plays the role of a cosmological four-vector necessary in this nonconservative structure. Henceforth we work in the units $c=1$ and $8 \pi G =1$.

For the purposes of our study we are going to use the conformal Newtonian gauge, whose metric is given by
\begin{equation}
ds^{2}=a(\eta)^{2}\left[-(1+2\Phi)d\eta^{2}+(1-2\Psi)\delta_{ij}dx^i dx^j\right].
\label{ds2}
\end{equation}
Since in (\ref{ds2}) the time evolution is parametrized by the conformal time, the four vector $\lambda_{\mu}$ shall be rewritten as $\lambda_{\mu}\rightarrow \tilde{\lambda}_{\mu}$, where 
\begin{equation}
\label{lmu}
\tilde{\lambda}_{\mu}= \left(a \lambda_0, \vec{0}\right).
\end{equation}
This nonconservative theory is sourced by the energy momentum of a perfect fluid
\begin{equation}
T^{\mu}_{\nu}=(\rho+p)u^{\mu}u_{\nu}+p\delta^{\mu}_{\nu},
\end{equation}
where the four-velocity obeys the constraint $u_{\mu}u^{\mu}=-1$.

Solving $K^{\mu}_{\nu}$ for the metric (\ref{ds2}), one finds the following components
\begin{eqnarray}
K^{0}_{0}=\frac{3 \mathcal{H} \lambda_{0}}{a} - \frac{3\lambda_0}{a}\Psi^{\prime}-\frac{6 \lambda_0 \mathcal{H} }{a}\Phi,\\
K^{i}_{j}= \frac{\lambda_0}{a} \left[ \mathcal{H} -2\mathcal{H} \Phi - \Psi^{\prime} \right]\delta^{i}_{j}, \\ 
K^{i}_{0}=\frac{2\lambda_0}{a}(\partial_{j}\Phi)\delta^{ij}, \\
K_{i}^{0}=-\frac{2\lambda_0}{a}(\partial_{i}\Phi), \\
K= \frac{6 \mathcal{H}}{a}\lambda_0 -\frac{12\lambda_0 \mathcal{H}}{a}\Phi-\frac{6\lambda_{0}\Psi'}{a},
\end{eqnarray}
where $\mathcal{H}=a^{\prime}/a$ and the prime symbol means derivative with respect to the conformal time. Aiming at analyzing separately the both regimes, let us now extract from the above equations both the background and the perturbative contributions.

\subsection{Cosmological background}

Let us begin by considering a universe with one single matter fluid, whose background expansion is fully described by the equations below

\begin{eqnarray}
\mathcal{H}^{2}=\frac{a^2 \rho}{3}, \\
2 \mathcal{H}^{\prime}+\mathcal{H}^2+2a \mathcal{H}\lambda_0=- a^2 p,\\
\rho^{\prime}+3\mathcal{H} \left(\rho + p +\frac{2 \lambda_0 \mathcal{H}}{a}\right)=0.
\label{cont}
\end{eqnarray}

It is worth noting that at the background expansion level the modifications induced by the tensor $K_{\mu\nu}$ is equivalent to the usual GR description sourced by a fluid with effective pressure 
\begin{equation}
p_{eff}= p + \frac{2 \lambda_0 \mathcal{H}}{a},
\end{equation}
which resembles the same structure as a bulk viscous fluid pressure. This means that regarding $\lambda_0 <0$ accelerated expansion is potentially achieved.

Since we are focusing on the late-time aspects of the Universe, let us neglect the radiation contribution to the cosmological expansion. We assume a universe purely dominated by a dark matter endowed with a pressure $p$ and an energy density $\rho_m$ which relates each other by means of an equation of state parameter given by $w\equiv p/\rho_{m}$. In the case of a one-fluid approach, i.e., ($\mathcal{H}\equiv \mathcal{H}(\rho_m)$), it is worth noting that the modified continuity equation (\ref{cont}) admits an analytical expression for the matter density evolution leading to
\begin{equation}
\frac{\mathcal{H}}{\mathcal{H}_0}=-\frac{2 \lambda_0}{3\mathcal{H}_0(1+w)a}+\left[1+\frac{2\lambda_0}{3 \mathcal{H}_0(1+w)}\right]a^{-(5+3w)/2}.
\label{Hlambda}
\end{equation}
From this equation we can calculate the deceleration parameter. Its current value is 
\begin{equation}
q_0=-\left[1-\frac{3(1+w)}{2}-\frac{\lambda_0}{\mathcal{H}_0}\right]. 
\label{q0}
\end{equation}
From the above expression, the current phase of accelerated expansion is achieved only if $\lambda_0/\mathcal{H}_0 < -(0.5+1.5 w)$. Considering that, throughout this work, we shall always stick to the cases when $w$ is very close to zero ($w \sim 10^{-7}$ at most), it is reasonable to set $\lambda_0/H_0 \lesssim -0.5$ as the upper bound to be respected by the main parameter of the model. From now on, whenever we refer to the today's Hubble factor, we shall use $H_0$ (i.e, defined in terms of the cosmic time) instead of $\mathcal{H}_0$. The main reason is a more clear physical meaning which is present in the former when comparing with the latter. However, notice that the results we have obtained in the upcoming sections are not spoiled by this choice, since $H_0 = \mathcal{H}_0$.

\subsection{Cosmological perturbations}

We can compute now the components of the field equations. Considering absence of shear we are led to fix $\Psi=\Phi$. Taking this assumption into account we can write the relevant ones for our study, namely the ($0-0$) and ($i-j$), given by
\begin{eqnarray}
\nabla^{2}\Psi - 3 \mathcal{H}\Psi^{\prime}-3 \mathcal{H}^2 \Psi =  \delta \rho  / 2 ,
\label{00pert}
\end{eqnarray}
and 
\begin{eqnarray}
\Psi^{\prime\prime}+( 3 \mathcal{H} +a\lambda_0)\Psi^{\prime}+ (2 \mathcal{H}^{\prime} + \mathcal{H}^2 +  2a \mathcal{H}\lambda_0)\Psi=\frac{a^{2}\delta p}{2},
\label{ijpert}
\end{eqnarray}
respectively. Here we have set $\frac{8 \pi G}{c^4}=1$. Our strategy is to solve numerically Eq. (\ref{ijpert}) for the potential $\Psi$ and use its solution in Eq. (\ref{00pert}) to obtain the growth of matter perturbations $\delta = \delta \rho / \rho$. Contrary to the bulk viscous cosmology, the $\lambda_0$ contributions do not add scale-dependent terms to the linear dynamics.

An alternative procedure is to manipulate the conservation law (\ref{consL}). The only nontrivial equations emerging from it are the first-order continuity and Euler equations, namely
\begin{equation}
\label{cont1}
(\delta \rho)'+\rho(1+w) \theta+3 \mathcal{H}\delta \rho(1+c^{2}_{s})-3\Psi'\rho(1+w)=\frac{12\lambda_0}{a}(\mathcal{H}\Psi'+\mathcal{H}^2\Psi)-\frac{2\lambda_0}{a}\nabla^2 \Psi
\end{equation}
and
\begin{equation}
\label{euler} 
\rho (1+w)\theta'+\rho'(1+w)\theta + c^{2}_{s} \nabla^2(\delta \rho)+4 \mathcal{H}\rho (1+w) \theta + \rho (1+w)\nabla^2 \Psi=-4 \lambda_0 \mathcal{H} \nabla^2 \Psi,
\end{equation}
with $\theta\equiv \partial_{i}v$, $v^{i}=v_{i}\equiv v$ and $c_{s}^{2}\equiv \delta p/\delta \rho$ defines the sound speed. In the equations above we can use the background continuity equation (\ref{cont1}) to get rid of the $\rho'$ terms. Besides, it is convenient to study the system in the quasi-static regime for which the spatial derivatives dominate over the temporal ones. Considering such a hypothesis, we can combine (\ref{cont}) and (\ref{euler}), and take the time derivative of the resulting expression to describe the behavior of the linear density perturbations by means of the single second-order differential equation below:
\begin{eqnarray}
\label{2nd}
&&\delta''+\left[\mathcal{H}-3\lambda_0 a - 3 \mathcal{H}(2w-c_{s}^{2})\right] \delta'-\left\{\frac{3\mathcal{H}^2}{2}\left[1+w+(1-3w)(w-c_{s}^{2})\right]-3 a\lambda_0 \mathcal{H}(w-c_{s}^{2})\right.\nonumber\\ &-&\left. 9\mathcal{H}w\left[\lambda_0 a+\mathcal{H} (w-c_{s}^{2})\right]+4\lambda_0 a \mathcal{H}\left(1+\frac{3}{2}c_{s}^{2}\right)-2 (\lambda_0 a)^2 \right\} \delta-c_{s}^{2}\nabla^2\delta=0.
\end{eqnarray}
The above equation has to be solved numerically in order to assess the evolution of the density contrast and consequently the growth function according to
\begin{equation}
D(a)=\frac{\delta(a)}{\delta(a_0)} \rightarrow f(a)\equiv \frac{d\, {\rm ln} \,D(a)}{ d\,{\rm ln} \,a}.
\end{equation}
Today's scale factor is set to unity, $a_0=1$. The variance of the density field smoothed on $8 h^{-1} Mpc$ scales varies in time linearly with the normalized growth such that $\sigma^{2}_8(a)=\sigma^{2}_8(a_0) D(a)$. We assume for this value of the variance of the density field at $a_0$ the value $\sigma^{2}_{8}(a_0)=0.8$, which is consistent with current observations.

\begin{figure*}
\centering
\includegraphics[width=0.4\textwidth]{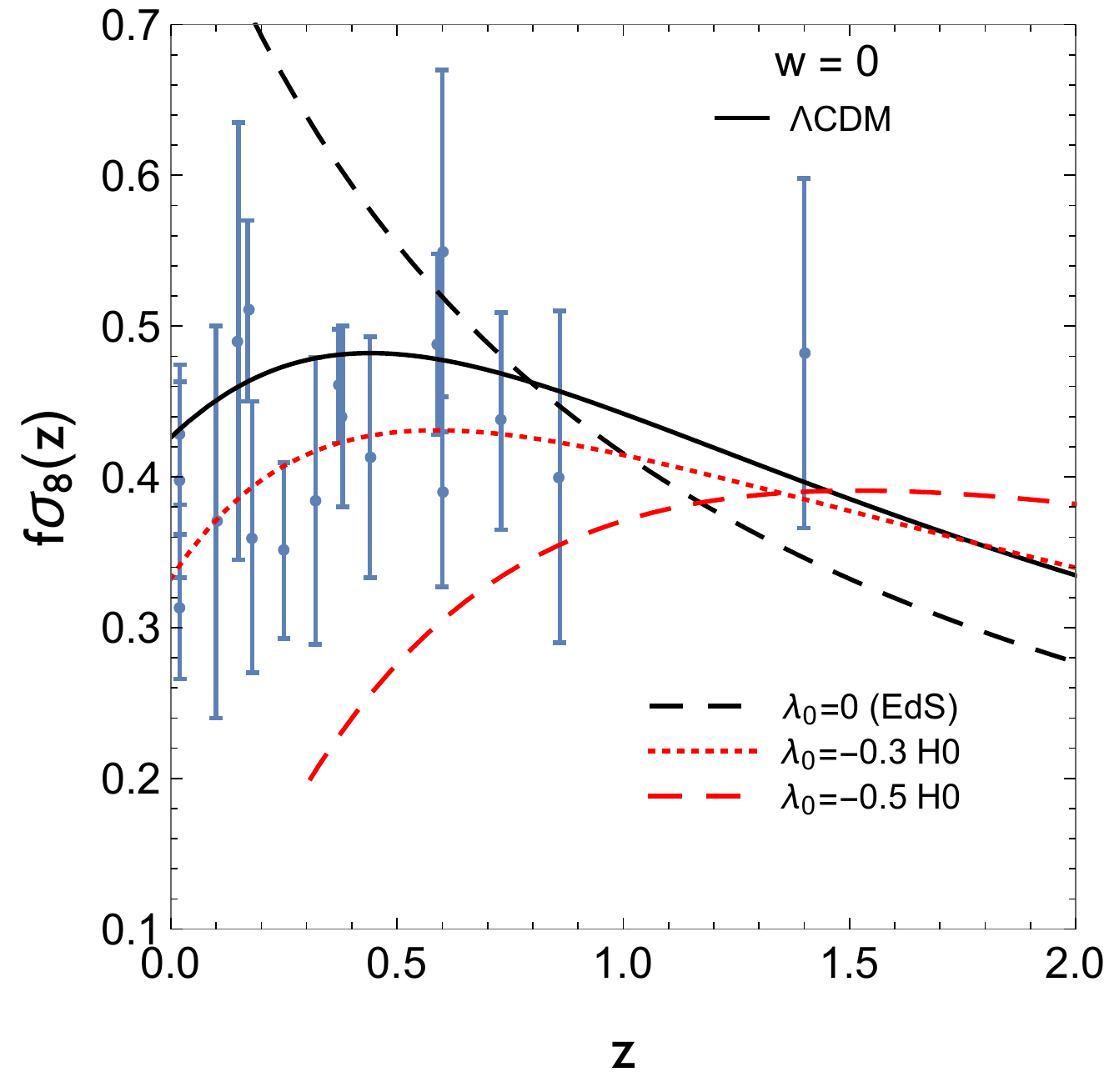}
\includegraphics[width=0.4\textwidth]{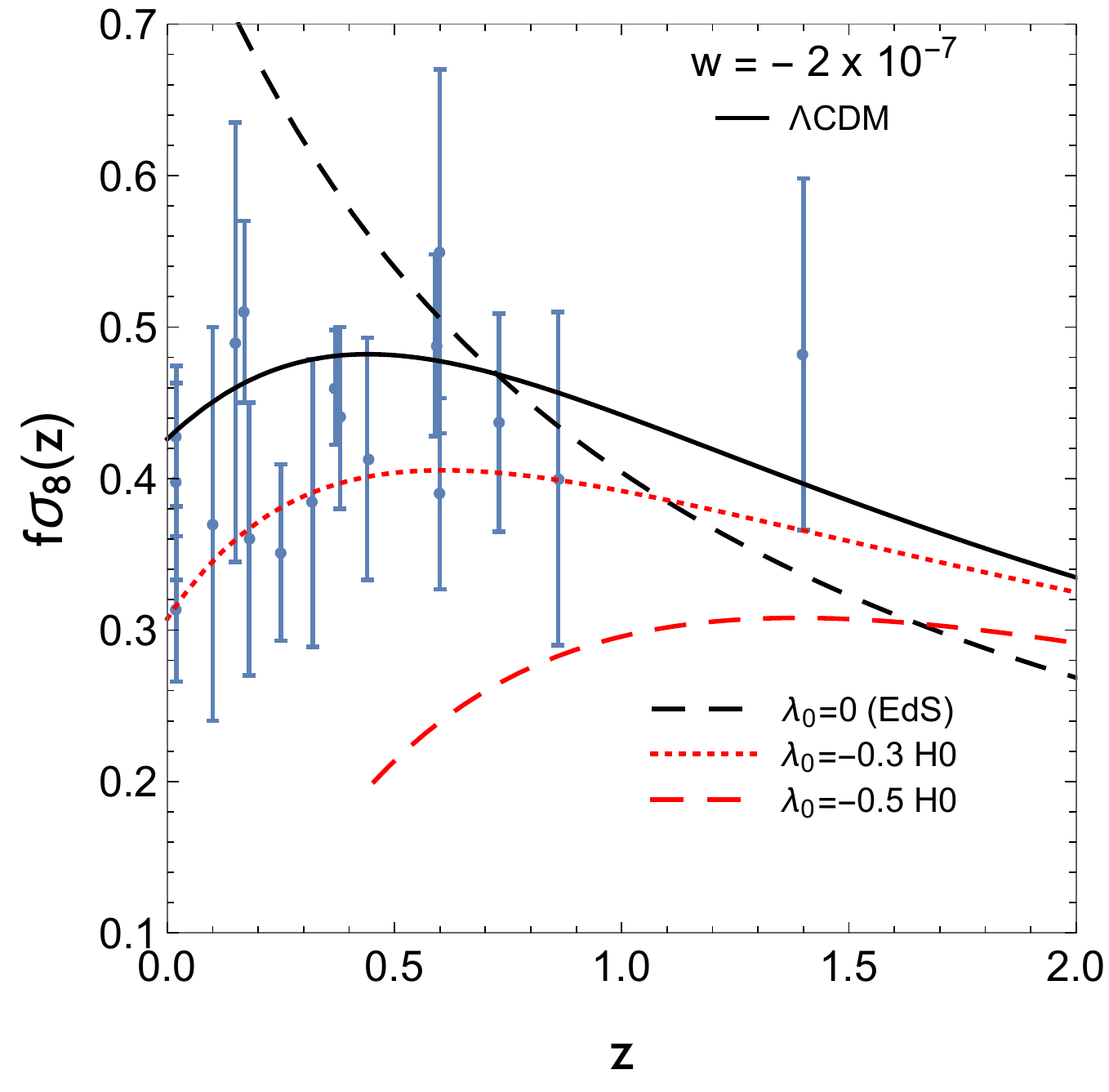}
\includegraphics[width=0.4\textwidth]{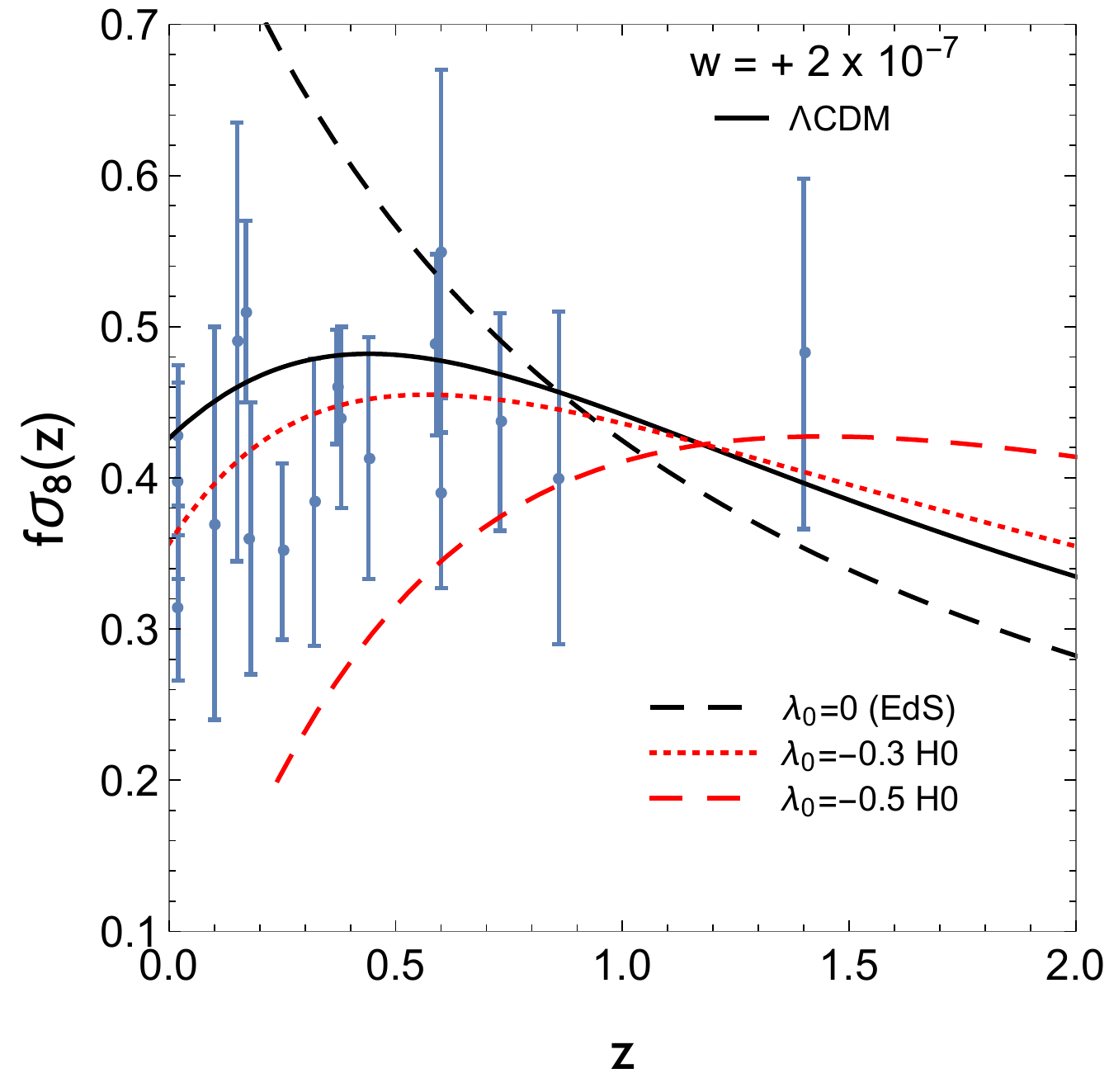}
\includegraphics[width=0.4\textwidth]{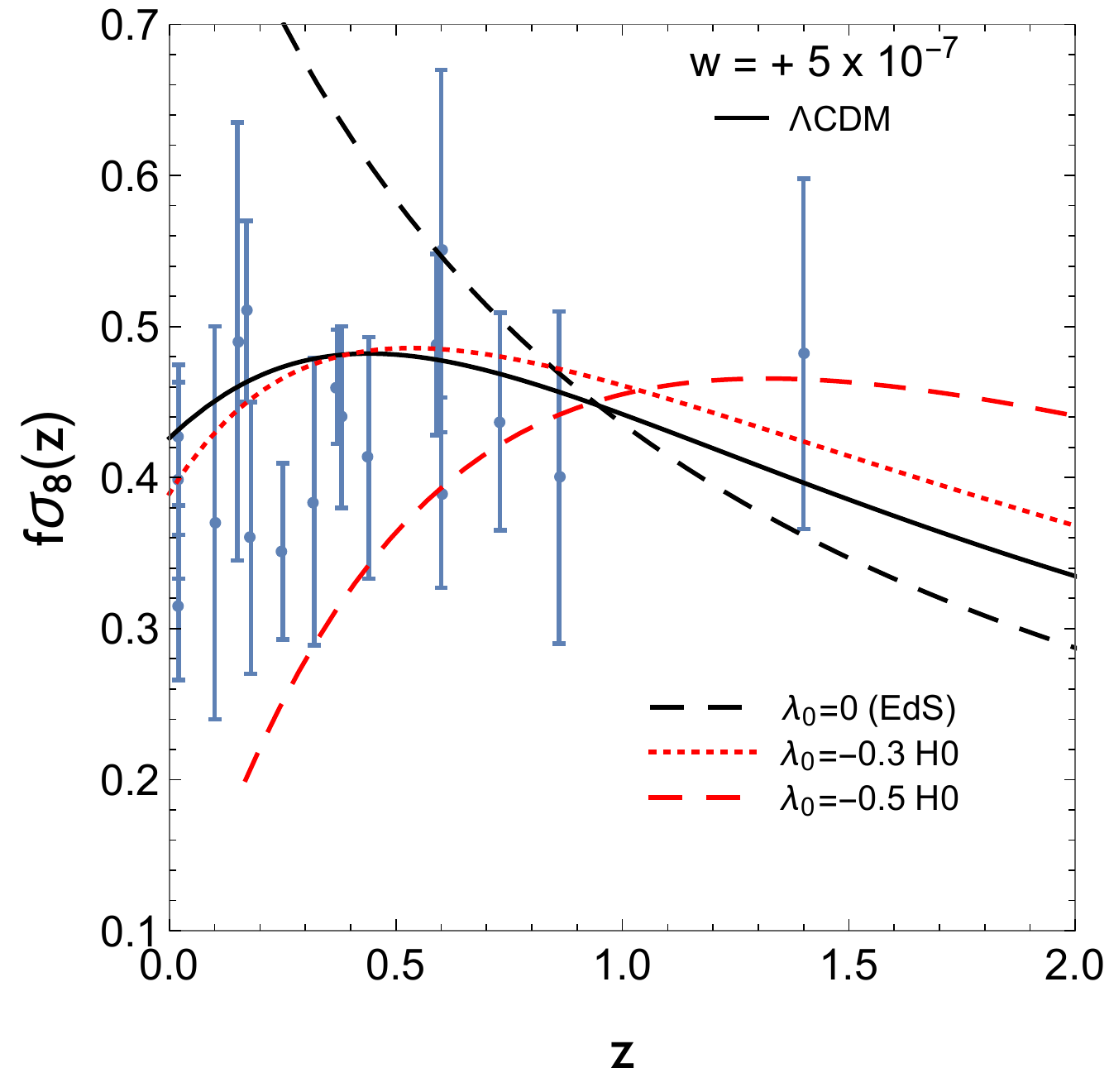}
\caption{The evolution of $f \sigma_8$ as a function of the redshift. The solid black line represents the standard $\Lambda$CDM model and the dashed black line the Einstein-de Sitter model. Dotted (dashed) red line is plotted for the nonconservative model with $\lambda_0 = -0.3 H_0$ ($\lambda_0 = -0.5 H_0$). Accelerating cosmologies occurs only for $\lambda_0 < -0.5 H_0$.}
\label{Fig1}
\end{figure*}

In Fig. \ref{Fig1}, we present results for the $f \sigma_8$ observable using the model with expansion rate (\ref{Hlambda}) with perturbations given by (\ref{2nd}). Observational data presented in all figures is the GOLD RSD sample compiled Ref. \cite{GOLD}. The solid black line represents the standard flat $\Lambda$CDM model. The dashed black line corresponds to the Einstein-de Sitter cosmological model which is the same as adopting $\lambda_0=0$. In the left-top panel of Fig. \ref{Fig1} we assume the Universe is filled with pressureless matter ($w=0$). The dotted red line $\lambda_0 = -0.3 H_0$ seems to a reasonable fit to the data. However, at the background level this value does not provide an accelerated expansion. According to Eq. (\ref{q0}) accelerating cosmologies are reached only if $\lambda_0 < -0.5 H_0$. Such values (see the dashed red line), on the other hand, clearly do not lead to acceptable data fitting.  

One can wonder whether the equation of state parameter of the matter fluid is able to modify this picture. As shown in the remaining panels of Fig. \ref{Fig1}fmatter  where we have adopted $w \neq 0$ values. This parameter impacts the evolution of the matter perturbations since it is proportional to the speed of sound of the fluid. As seen in Fig. \ref{Fig1} values or order $w \sim 10^{-7}$ impact the growth function but they lead to no significant changes to the background. Therefore, the parameter $w$ is not able to heal this pathological behavior. 

This fact reveals another similarity of this model with the case of a unified bulk viscous model in which the set of available model parameter space that fits the background expansion is not in agreement with first-order observables like the matter power spectrum or the CMB data \cite{Velten:2011bg, Velten:2012uv}.

\section{Pressureless dark energy model}
In the above section, we have realized that a one-fluid description of the cosmic substratum does not represent a viable model in the present nonconservative scenario. Then we propose now a new strategy considering a two-fluid model where the matter sector is decomposed into two pressureless components as follows  
\begin{equation}
\label{tot}
T^{\mu\nu} \rightarrow T^{\mu\nu}_{m}+T^{\mu\nu}_{x}.
\end{equation}
The component $T^{\mu\nu}_{m}$ is conserved as usual, whereas the remaining component, denoted by $T^{\mu\nu}_{x}$, obeys the modified conservation law (\ref{consL}). At some extent this decomposition can be interpreted as the dark sector of the Universe being composed by two types of pressureless dark matter. However, one of them couples to the geometrical sector via the nonconservation appearing in the theory described above. At first sight, this model does not possess any special advantage with respect to the standard $\Lambda$CDM, as it is also endowed with two dark fluids of an unknown nature. However, notice that contrary to the standard cosmology, the present model carries a dark energy component obeying a dust equation of state instead of a vacuum one. This feature makes such a model ``less exotic" than $\Lambda$CDM, since the dark energy fluid by itself does not lead to the violation of any energy conditions and also allows for the avoidance of the so-called discrepancy between observed and the theoretically predicted results for the vacuum energy density. The assumptions above imply in 
\begin{equation}
\rho^{\prime}_m+3\mathcal{H}\rho_m=0,
\label{consm}
\end{equation}
and
\begin{equation}
\rho^{\prime}_{x}+3 \mathcal{H} \left(\rho_x+\frac{2 \lambda_0 \mathcal{H}}{a}\right) + \rho^{\prime}_{m}+3\mathcal{H}\rho_m=0.
\end{equation}

Since we have adopted the usual conservation of the matter component (\ref{consm}) then
\begin{equation}
\rho^{\prime}_{x}+3 \mathcal{H} \left(\rho_x+\frac{2 \lambda_0 \mathcal{H}}{a}\right)=0
\label{rhox}
\end{equation}
The expansion rate now also depends on $\rho_m$
\begin{equation}
\mathcal{H}^{2}=\frac{a^2}{3} (\rho_m + \rho_x).
\label{Htwo}
\end{equation}
Therefore, equation (\ref{rhox}) does not admit trivial analytical solution. Rather we obtain the evolution of $\rho_x$ by inserting (\ref{Htwo}) into (\ref{rhox}). 

\begin{figure*}
\centering
\includegraphics[width=0.4\textwidth]{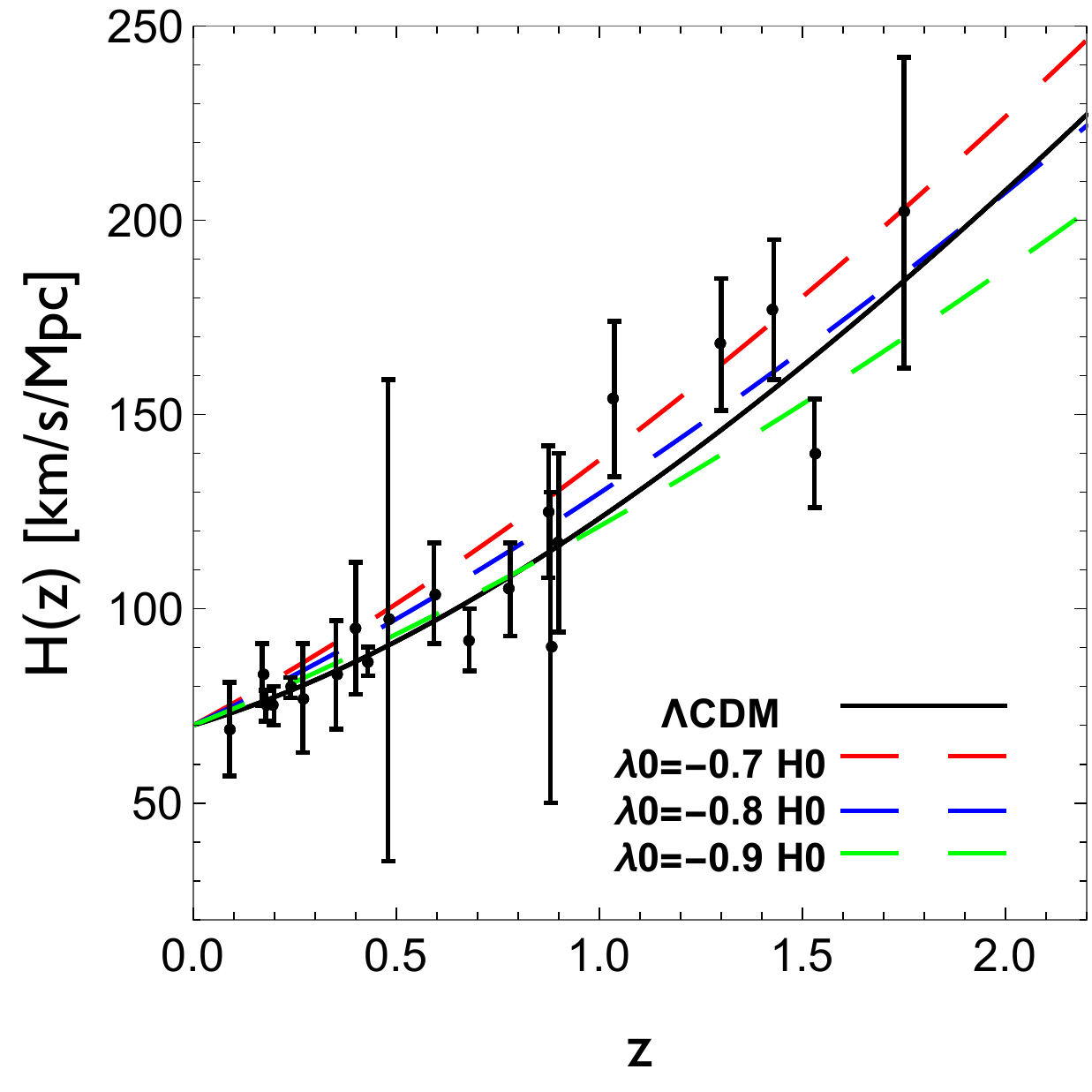}
\includegraphics[width=0.4\textwidth]{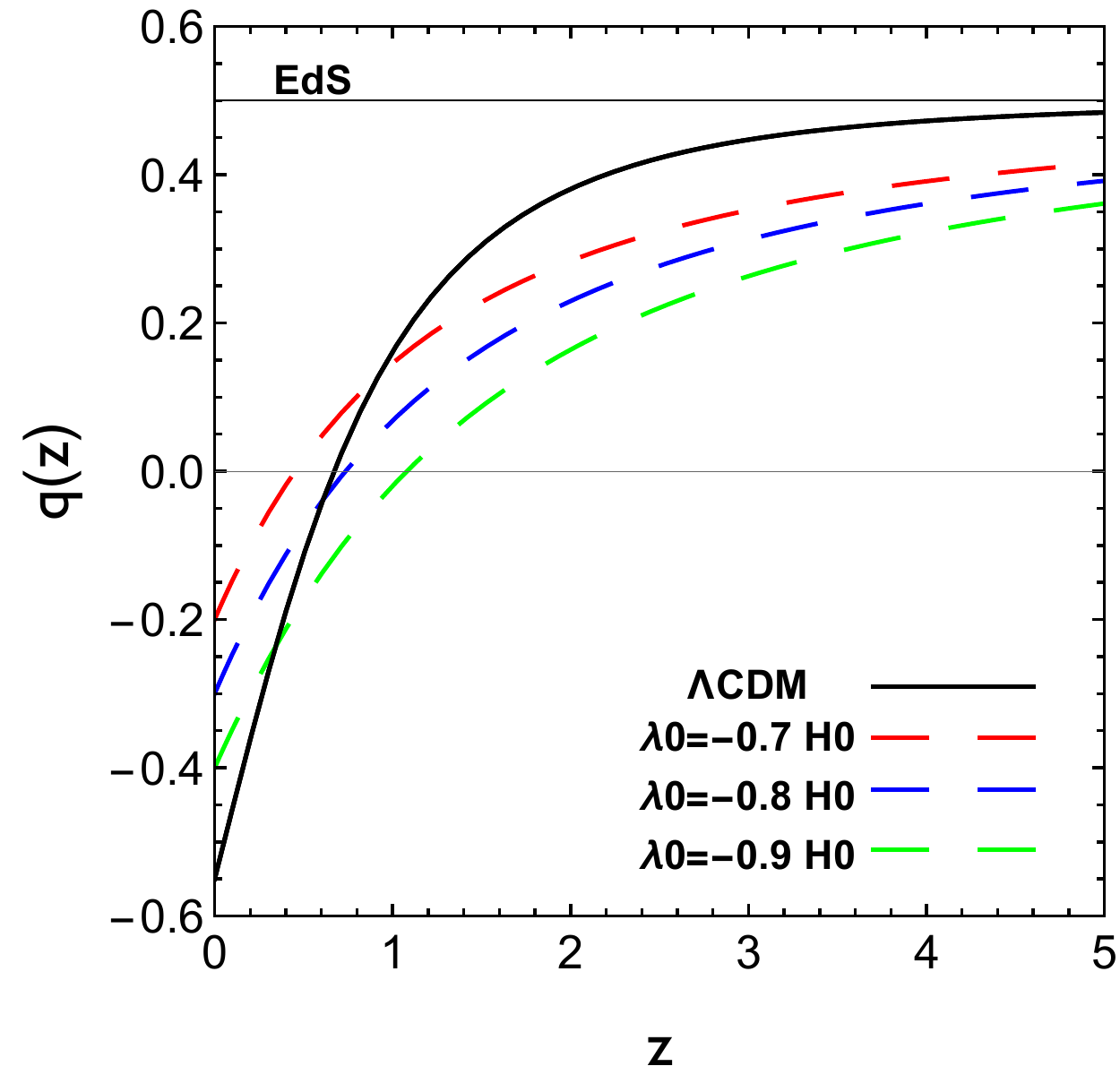}
\caption{ Evolution of background quantities for the two-component model.}
\label{Figtwo}
\end{figure*}

The background expansion rate and the deceleration parameter for the double dark matter model are plotted in Fig. \ref{Figtwo}. It is worth noting that fitting provided by the $\lambda_0$ values shown is reasonably consistent with the $H(z)$ data obtained from Ref.
\cite{Farooq:2016zwm}. 

Rather than performing a full statistical analysis to determine the best-fit parameter our aim now is to show that admissible values of the $\lambda_0$ at the background level can also lead to good fit of the perturbed quantities. In order to analyze this issue we calculate the scalar density perturbations of the conserved fluid ($m$) which obey to the equation
\begin{equation}
\delta_m^{\prime\prime} + \mathcal{H}\delta_m^{\prime} - \nabla^2 \Psi=0.
\label{pertdeltam}
\end{equation}
The gravitational potential $\Psi$ drives the evolution of matter fluctuations $\delta_{m}$. Our strategy now is to solve equation (\ref{ijpert}) for the case $\delta p =0$ since we have assumed both fluids are pressureless. We obtain then the evolution for $\Psi$ which can be inserted into (\ref{pertdeltam}). Finally, Eq. (\ref{pertdeltam}) is ready to be solved for the matter density contrast $\delta_m$. 

\begin{figure*}
\centering
\includegraphics[width=0.4\textwidth]{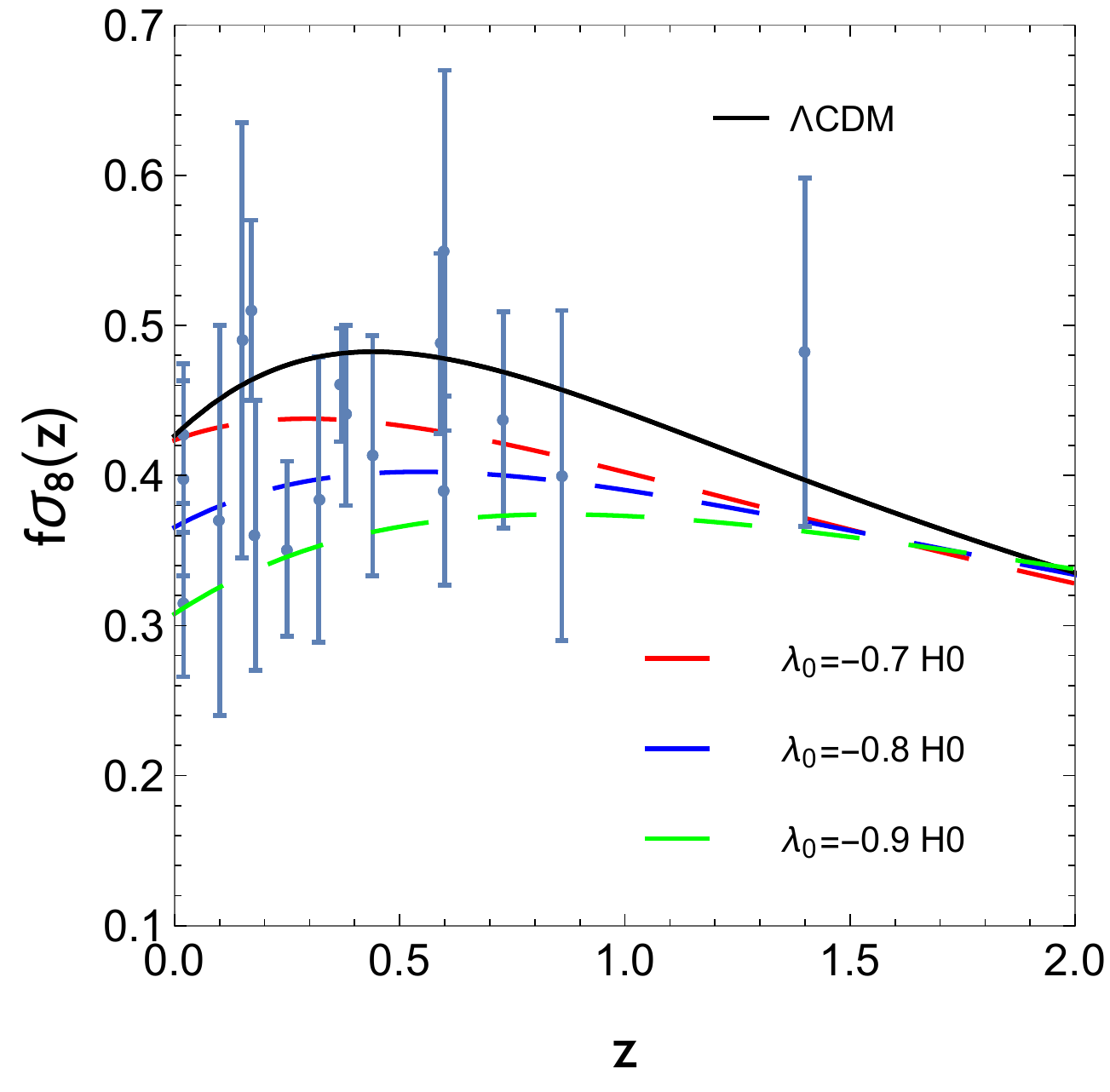}
\caption{ Impact of $\lambda_0$ on the evolution of the $ f \sigma_8$ observable for the two-component model. In this case, $ f \sigma_8$ is that one obtained from the $\rho_m$ component which obeys separately the usual conservation law.}
\label{Fig3}
\end{figure*}

Fig. \ref{Fig3} shows the $f\sigma_8$ observable corresponding to the same $\lambda_0$ values used in Fig. \ref{Figtwo}. Now, parameter values in the interval $-0.9 H_0< \lambda_0<-0.7 H_0$ seem to yield to a viable description of available data. Indeed, only a full statistical analysis would provide the best fit parameters and indicate whether or not this model can be competitive to the $\Lambda$CDM model, but this is beyond the proposal of this work.

\section{Conclusions}

In this work we have explored the cosmological consequences of the idea of action-dependent Lagrangians. Although this idea relies in the realm of nonstandard approaches for covariant theories of gravity, it has been deeply analyzed in the recent literature.

By constructing the cosmological solutions of a FLRW expansion and the scalar perturbations around it we have demonstrated in this work what kind of cosmologies appear in this scenario.

In particular, by sourcing the resulting field equations with a single fluid which due to the intrinsic features of the theory does not obey the usual conservation law, it is worth noting that the effective dynamics resembles that one of a bulk viscous fluid. As widely explored in the literature and also shown here in Sec. II such proposal of a single component driven the FLRW expansion does not provide a viable description of observational data.

In Sec. III we introduced the strategy of splitting the total energy momentum tensor into two pressureless components. One of then does not couple to the geometric sector of the theory while the other one does. The former behaves therefore as a typical cold dark matter fluid while the latter plays the role of an effective dark energy  fluid yielding to a consistent accelerated expansion at late times. From the astroparticle point of view our viable model can
be composed by two distinct dark matter-like particles. Also, by analyzing the growth of scalar perturbations in such double pressureless components we find a reasonable agreement with available data. 

Our main goal in this work was to set up the a viable cosmological model composed by two pressureless fluids, one of them coupled to the geometrical sector via the features imposed by the action-dependent Lagrangian formalism. Though we have provided here only the qualitative aspects of the model but demonstrated its viability, we hope that a full statistical analysis with an enlarged set of observational data will provide the best-fit parameters of this model. We hope to present these results in a future work.

\noindent
{\bf Acknowledgements:} Partial financial support by FAPES, CNPq and CAPES (Brazil) is gratefully acknowledged.

\end{document}